\mag=\magstephalf
\pageno=1
\input amstex
\documentstyle{amsppt}
\TagsOnRight
\interlinepenalty=1000
\hsize=6.4truein
\voffset=23pt
\vsize =9.8truein
\advance\vsize by -\voffset
\nologo

\NoBlackBoxes
\font\twobf=cmbx12
\define \ii{\roman i}
\define \ee{\roman e}
\define \dd{\roman d}
\define \cn{\roman {cn}}

\define \am{\roman {am}}

\define \cls{{\roman{cls}}}
\define \qcl{{\roman{qcl}}}
\define \reg{{\roman{reg}}}

\define \Em{{\roman{Em}}}

\define \period{{\roman{period}}}
\define \traj{{\roman{traj}}}
\define \GL{{\roman{GL}}}
\define \PSL{{\roman{PSL}}}
\define \SSp{{\roman{Sp}}}
\define \tvskip{\vskip 0.5 cm}

{\centerline{\bf{Statistical Mechanics of Elastica on Plane as a Model of 
Supercoiled DNA}}}
{\centerline{\bf{------Origin of the MKdV hierarchy-------}}}
\author
\endauthor
\affil
Shigeki MATSUTANI\\
2-4-11 Sairenji, Niihama, Ehime, 792 Japan \\
\endaffil
\endtopmatter

\subheading{Abstract}

In this article, I have investigated statistical mechanics of a
 non-stretched elastica in two dimensional space using  path 
integral method.  In the calculation,  the MKdV hierarchy 
naturally appeared as the equations including the temperature 
fluctuation.I have classified the moduli of the closed elastica
 in heat bath and summed the Boltzmann weight with the 
thermalfluctuation over the moduli. Due to the bilinearity of the
 energy functional,I have obtained its exact partition function.By
 investigation of the system,I conjectured that an expectation 
value at a critical point of this system obeys  the Painlev\'e 
equation of the first kind and its related equations extended 
by the KdV hierarchy.Furthermore I also commented onthe relation
 between the MKdV hierarchy and BRS transformationin this system. 

\document


\tvskip
\centerline{\twobf \S 1. Introduction }
\tvskip

Elastica has sometimes appeared in the history of mathematical physics
 according to refs.[1-4].
The problem of elastica,  an ideal thin elastic rod, was proposed by 
James Bernoulli in 1691.
By investigation on the behavior of an elastica,  Bernoulli's family and 
their related people, Euler, d'Alembert and so on,
discovered many non-trivial mathematical and physical facts, {\it e.g.},
classical field theory, minimal principle, elliptic function, mode analysis,
 non-linear differential equation and others [1-4]. 
In fact, James Bernoulli derived the elliptic integral  related to the 
lemniscate function in 1694, before Fagnano found his lemniscate function 
[1-3], and found that the force of elastica is proportional to inverse of 
its curvature radius [1]. His nephew, Daniel Bernoulli followed James's  
discoveries and discovered the energy functional of 
elastica and its minimal principle around 1738. Succeeding Daniel's 
and James's discoveries, Euler derived  elliptic integral of 
general modulus as a shape of elastica using Daniel's minimal 
principle and  numerically integrated it. Then he completely classified 
shapes of 
static elasticas by numerical sketch [1], which are, nowaday, 
known as special cases of loop soliton [5].
In the computations, Euler used the static sine-Gordon equation. 
These computations  essentially imply discovery of the integrable nonlinear 
differential equation, the investigation of its moduli
 and first application of  algebro-geometrical functions to physics.
It should be noted that they came from the studies on the elastica.
Furthermore it is well-known that the elastica problem is the simplest 
 prototype $\sigma$-model $S^1\to \text{SO}(2)$ or SO(3) [6,7],
  which was found in 18th century and investigated by Kirchhoff in last 
  century [4].

Thus the elastica problem sounds to be legacy before last century but
 I believe that its properties are not completely understood and
its role in mathematical physics is still important.
The difficulty to solve the elastica problem is that one must consider
 the constraint condition such as the isometry (non-stretch) condition 
 and boundary condition.
In fact, even though it is not an elastica problem,
Goldstein and Petrich naturally rediscovered the MKdV hierarchy
through (virtual) dynamics of a space curve with isometry condition [8,9]. 
(Readers should noted that  the time-development of the physical 
elastic rod is not governed by the MKdV hierarchy in general [7,9-14] 
even though in  some papers it seems to be misunderstood.)
Due to their work, the MKdV hierarchy can be geometrically realized.
Furthermore, using their construction, I proved that the Hirota bilinear
equation  and $\tau$ function of MKdV hierarchy [15]
can also be translated into  the geometrical language of a space curve [11]. 
The classical analogue of the vertex operator  naturally 
appears as a complex tangent vector of the elastica [11].

However these approaches are just mathematical and geometrical ones but
are not directly connected with physical problem because in most of works 
there is little physical reason
why the MKdV hierarchy must appear in physical system [8,9].

On the other hand, the study of elastic chain model of 
deoxyribonucleic acid (DNA) is current [16-22].
DNA usually occurs as a double helix with two complementary
nucleotide chains winding around a common axis. In cells,
the common axis of the looped  DNA often folds into intricate structure, 
or a supercoiled form [16,17]. Due to the enormous size of DNA,
conformation of double helical polymer needs topological and 
geometrical studies [23]. Thus the double helical polymer is often 
modeled as a thin elastic isotropic rod or an elastica and studied from
the kinematic and topological point of view.
Topological invariance such as the linking number 
classifies the shape of the DNA in three-dimensional euclidean space 
$\Bbb R^3$ [16].
Based on the Kirchhoff's model of an elastica in $\Bbb R^3$
and the nonlinear Schr\"odinger equation,
the possible kinematic conformations of DNA were considered [16-21].
Partial thermal effect on their shapes were argued in ref.[18].
Furthermore the molecular dynamics study combining with the elastic
energy model was reported [22] and then an topology change related 
to the knot was found.

Statistical mechanics of elastic chain as a model of a large polymer 
like DNA was studied by Sait\^o {\it et al.} using the path integral [24].
They calculated some exact partition function of the energy functional 
of elastic chain.
Such chain model is sometimes called wormlike chain model [25].
Their approach is successive in the polymer physics and influences recent 
works [26 and references therein]. However they did not pay any attention 
upon  isometry condition in a calculation of the path integration
even though they imposed isometry condition after computation 
(a kind of quantization);
they summed  allover configuration space without isometry condition.
It should be note that the constraint does not commute with quantization 
and statistical treatment [27].
(An example of such inconsistency is shown in ref.[28].)

It is a natural assumption that flexible polymers which cannot  stretch as 
a classical object (at zero temperature) cannot stretch even in heat bath. 
Hence it is very important to calculate the partition function
of the elastica with isometry condition and sum  the Boltzmann weight over 
only allowed conformations. 

The purpose of this article is to investigate the behavior of a closed 
large polymer like DNA in heat bath of two-dimensional space, 
whose shape is determined by its elasticity and stretch is negligible, 
and to clarify the physical meaning of the MKdV hierarchy.

Furthermore it is known that DNA sometimes exhibits topology changes and 
has topological isomer [22,29]. In this article, as I will deal with an 
elastica in two-dimensional space, there is no knot invariance but exists
writhing number as a topological invariant of the elastica in 
two-dimensional space if  crossings are allowed [28].
Thus in this article, even though the elastica in two-dimensional
space will be investigated, I will allow the crossings if it can be 
realized when I embed it in three-dimensional space. Then I will  
argue the topological change on this problem.

The organization of this article is as follows.
\S2 reviews classical shape of an elastica in two-dimensional 
flat space adding infinity point, {\it i.e.}, 
$\Bbb C P^1$. In \S 3, I will investigate the statistical mechanics 
of an elastica.
First, I will consider the thermal fluctuation of the extremum point of a 
partition function of the elastica in
terms of the path integral method. Then I will obtain the MKdV hierarchy 
by physical requirements. Second, I will investigate the moduli space 
of the quasi-classical elastica, which is the extremum point of 
the partition function. Finally I sum the Boltzmann weight over the moduli 
space and obtain an exact formulation of the partition function.
In \S4, I will discuss the obtained partition function and I will comment 
upon the relation between the Goldstein-Petrich method  and the BRS 
quantization of the gauge field [31] and a critical point of this model.

\tvskip
\centerline{\twobf \S 2. Classical Shape of   Elastica}
\tvskip

Here I will quickly review a shape of a closed elastica in two-dimensional 
space [10-12,32]. I will denote by $C$ a shape of the elastica  embedded 
in the projective complex line (or the Riemann sphere) $\Bbb C P^1$
and by $X(s)$ its affine vector : 
$$
	S^1 \ni s \mapsto X(s) \in C \subset \Bbb C P^1 ,
	\quad X(s+L)=X(s),
	\tag 2-1
$$
where $L$ is the length of the elastica.
The Frenet-Serret relation will be expressed as
$$
	\exp(\ii \phi)=\partial_s X, \quad
	\partial_s \exp(\ii \phi)= \ii k \exp(\ii \phi)  ,\tag 2-2
$$
where $\phi$ is a real valued function of $s$ and $k$ is the curvature of
the curve $C$, $k:= \partial_s \phi$: $\phi(s+L)=\phi(s)$ and $k(s+L)=k(s)$.
 Here I have chosen  the metric of the curve as the induced metric 
 from the natural metric of $\Bbb C \subset \Bbb C P^1$; by the choice
$\phi$ is real valued.

As Daniel Bernoulli suggested to Euler [1],  the energy integral of 
the elastica is given as
$$
	E=\int_0^L \dd s k^2 ,\tag 2-3
$$
and  shape of the elastica is realized as its stationary point.
Here I assume that the elastica does not stretch and preserves 
its infinitesimal length;
since  deformation of the elastica is regarded as one parameter 
transformation, the assumption implies that the transformation is 
isometric [10-13,32]. 
Thus I will refer this condition as isometry condition.

Following the minimal principle, I will derive the differential equation.
I will consider the variation of the curve $C \to C_\varepsilon$ 
under isometry condition [8-11],
$$
       X \to X_\varepsilon = X + \varepsilon (U_1 + \ii U_2)\exp(\ii \phi),
       \quad U_1(L)=U_1(0), \quad U_2(L)=U_2(0) \tag 2-4
$$
where $\varepsilon U_1$ and $\varepsilon U_2$ are  
infinitesimal real valued functions.

Since the infinitesimal length  of the curve is given as
$$
	\dd s^2= \dd \bar X \dd X= \partial_s \bar X \partial_s  X\dd s^2,
	 \tag 2-5
$$
corresponding length of the $C_\varepsilon$,
$$
	\split
          \dd \bar X_\varepsilon \dd X_\varepsilon
          &=\left(1  +\varepsilon\left((\partial_s-\ii k) U_1 - 
          \ii(\partial_s -\ii k) U_2 \right) \right)
          \left(1  +\varepsilon\left((\partial_s+\ii k) U_1 
          + \ii(\partial_s +\ii k) U_2 \right) \right)  \dd s^2\\
          &= \left(1  +2\varepsilon\left(\partial_s U_1 
          - k U_2 \right) \right) \dd s^2 +\Cal O(\varepsilon^2)
          \endsplit,
           \tag 2-6
$$
must be $\dd s^2$ modulo $\varepsilon^2$ due to the isometry condition. 
Hence I obtain the relation,
$$
        \partial_s U_1 = k U_2  .\tag 2-7
$$
The tangential angle of $C_\varepsilon$ is given as
$$
	\split
	\phi_\varepsilon&=\frac{1}{\ii} \log \partial_s X_\varepsilon\\
		&=\phi+\varepsilon(k+\partial_s k^{-1} \partial_s) U_1 .
		\endsplit \tag 2-8
$$
Its curvature is expressed as
$$
	\split
	k_\varepsilon & = \exp(-\ii \phi_\varepsilon)\partial_s^2 
	X_\varepsilon \\
		     &= k+\varepsilon \partial_s 
		     \left((k+\partial_s k^{-1} \partial_s) U_1\right).
		    \endsplit \tag 2-9
$$
Finally I obtain the variation of the energy functional
$$
             \int_0^L \dd s k_\varepsilon^2 = \int_0^L \dd s 
             \left( k^2 +\varepsilon 
             \left( \frac{1}{2}\partial_s(k^2) +\partial_s
                     \frac{\partial_s^2 k}{k}\right)U_1\right)
             +\Cal O(\varepsilon^2). \tag 2-10
$$
From the variational equation,
$$
	\frac{\delta E[k_\varepsilon]}{\delta (\varepsilon U_1)}=0, 
	\tag 2-11
$$
the non-linear differential equation is given as the equation of 
the shape of the static elastica,
$$
	\partial_s(\frac{1}{2} k^2  +\frac{\partial_s^2 k}{k})=0,  
	\tag 2-12-a
$$
and thus 
$$
	a_1 k +\frac{1}{2} k^3 +\partial_s^2 k=0,  \tag 2-12-b
$$
where $a_1$ is the integral constant. This equation is known 
as the static MKdV equation
in the soliton theory if derivate it by $s$ again.

First I will note that (2-12)'s are also equations of 
the energy functional [10,11],
$$
	E=\int_0^L \dd s \left( k^2  + A_1 \cos \phi +A_2 \right), 
	\tag 2-13
$$
where the second term means the constraint for the relative position of 
$X(0)-X(L)$ and the third one is for the total length $L$ [7,10,11,13].
The sufficiency of the third term is trivial. From (2-9),  
the second term becomes
$$	
	A_1\int \dd s \cos (\phi_\varepsilon)
	=A_1 \int \dd s \left(\cos (\phi)
	-\varepsilon \left(k U_1+\partial_s 
	\frac{\partial_s U_1}{k} \right)\sin(\phi) \right) 
	+\Cal O(\varepsilon^2)\tag 2-14
$$
and by the partial integration, the second term in rhs vanishes for any 
$U_1$. This vanishing occurs owing to the compatibility between
 the MKdV equation and the static sine-Gordon equation,

$$
	\partial_s^2 \phi+ A_1 \sin \phi=0 . \tag 2-15
$$
(2-15) comes from natural variation of $\phi$ of (2-13) [1,4,7,10,11,13].
Hence (2-12)'s can be also regarded as the stationary point of (2-13).
In fact it is known that solutions of (2-12) are in agreement with 
those of (2-15) 
as I will show later. It should also be noted that (2-13) is 
the $\sigma$-model with the topological term and
 was discovered in 18th century. In other words, the system of 
 the elastica can be regarded as SO(2)-principal bundle over 
 $S^1$ and the cosine term in (2-13) is a local version
of the fundamental group $\pi_1(S^1)=\Bbb Z$ [6]. 

Solutions of (2-12) are completely expressed by the elliptic functions.
Multiplying $\partial_s k$ and integrating $s$ [4,11,13,14,32,33],
I obtain the relation,
$$
    (\partial_s k)^2 = -
    \frac{1}{4}(k^4 - a_1 k^2+a_2)  .   \tag 2-16 
$$
Introducing the quantities, $\beta_2-\beta_2:=a_1$, 
$l= \sqrt{\beta_1/(\beta_1+\beta_2)}$, and 
$\beta_3 := \sqrt{\beta_1+\beta_2}/2$,  $k$ is expressed 
by the Jacobi elliptic function [34],  
$$
    k = \sqrt{\beta_1}\roman{cn}(\beta_3 s,l) .
                \tag 2-17
$$
Transformation from the solutions of (2-12) to those of (2-15) 
is given by the identities of the integrand in the elliptic integral 
between   trigonometric function and polynomial expressions [34].
After integrating the differential equations, I obtain [1,4,10,11,13,33],
$$
    X(s)= \frac{2}{\beta_3} E(\am(\beta_3 s),l) -s - \ii\frac{2 l }
    {\beta_3}( \cn (\beta_3 s)-1) ,  \tag 2-18
$$ 
where $E(\cdot,l)$ is the incomplete elliptic integral of the second kind
and $\am$ is the Jacobi amplitude function [34]. Due to the closed 
condition,
$$
X(0)=X(L),  \tag 2-19
$$ 
there is an eight-figure shape [1,4,11,13,16,33]; the modulus of the 
elliptic function is $l=0.908909\cdots$ and the ratio of the fundamental 
parameters is $K'/K=0.70946\cdots$.

Thus in the set of  solutions of (2.12), there are only two closed 
elasticas  in $\Bbb C$ up to translation of their centroid, global 
rotation and scaling;  circle $k=2 \pi/L$ and eight-figure shape.
Here though I have chosen  solutions such as $k=2 \pi /L$,
there also exist other solutions like $\{\ k=2 \pi n/L\ | \ n\ge 1 \}$.

On the other hand, by taking limit $L \to \infty$ and by considering ones 
in $\Bbb C P^1$, more various closed elasticas in $\Bbb C P^1$ are allowed.
These solutions were classified by Euler in 18th century [1,4]. 

In this article the set of elasticas obtained as solutions of (2-12) 
will be denoted as $\frak S_\cls$
$$
	\frak S_\cls=\{C | C\ \text{is a solution of (2-12)'s}\}, \tag 2-20
$$
and the energy functional is expressed by,
$$
	E_\cls[C] = \int_0^L \dd s\  k^2 , \quad C \in \frak S_\cls .
	\tag 2-21
$$

\tvskip
\centerline{\twobf \S 3. Statistical Mechanics of  Elastica}
\tvskip

In this section, I will consider statistical mechanics of 
a closed elastica and investigate its behavior at heat bath.
I continue to allow the crossing of the elastica even in two-dimensional 
space.
I set up that there are  many independent laboratory dishes in which  
large polymers like DNA individually exist  one by one.
A looped elastica is in the liquid whose temperature 
is determined and viscosity is very large. The liquid is a kind of heat 
bath. Then the kinetic energy of the elastica is suppressed in
equilibrium state due to dissipation  but owing to the fluctuation by 
temperature noise, the elastica sometimes jumps from a quasi-stable 
state to other quasi-stable states.

The partition function of the elastica is given as [24],
$$
	Z=\int D X \exp\left(-\beta \int^L_0 
	\dd s\left[ k^2\right] \right)  .\tag 3-1
$$
Here I go on to prohibit that  the local length of the elastica does 
 change; the isometry condition will be maintained.
In [24], it was written that if one also dealt with the kinematic term, 
it would be decoupled with the potential term (2-3).
However as I employ the isometry condition, this statement cannot be 
guaranteed at all because
the kinetic term is also restricted by the isometry condition [8,9]
and strongly coupled with the shape of the elastica.

\subheading{ Quasi-Classical Motion}

By the quasi-classical method in path integration [35,36], 
I will evaluate the partition function (3-1).
I will expand the affine vector around an extremum point of the integral,
$$
    X= X_{\qcl}+ \epsilon (u_1(s) + \ii u_2(s) )\exp(\ii \phi_{\qcl}) 
    + \Cal O(\epsilon^2),   \tag 3-2
$$
where $\epsilon$ is an infinitesimal parameter, 
$\epsilon \propto 1/\sqrt{\beta}$ and $\phi_{\qcl}$ is the tangent angle 
of the quasi-classical curve of elastica. $X_{\qcl}$ is an affine 
vector of the extremum point
of the functional integral and will be determined later. 
In the path integral, terms with higher orders of $\epsilon$ also play 
important role
and thus I must pay attentions upon the higher perturbations of 
$\epsilon$ here. Hence I will assume that $X$ is parametrized by a 
parameter $t$ and the difference between $X$ and  $X_{\qcl}$ 
can be expressed by,
$$
       X(s,t):=\ee^{\epsilon \partial_{t}}X_{\qcl}(s,t),
       \quad	\epsilon \partial_{t} X_{\qcl}  =X- X_{\qcl} 
       + \Cal O(\epsilon^2). \tag 3-3
$$
Since for an analytic function $f(x)$, $\ee^{a \partial_{x}}f(x)=f(x+a)$,
$X(s,t)$ can be expressed as $X(s,t)=X_{\qcl}(s,t+\epsilon)$, 
where direction $t$ differs from that of $s$; the domain of functional 
integration (3-1) deviates from the domain $S^1$ of the classical 
map (2-1). Then (3-2) becomes
$$
   \partial_{t} X_{\qcl} = (u_1 + \ii u_2 )\exp(\ii \phi_{\qcl}) , 
   \quad u_1(L)=u_1(0), \quad u_2(L)=u_2(0). 
     \tag 3-4
$$
This is virtual dynamics of the curve [8-10].
As well as the argument in \S 1, due to the isometry condition, 
I require $ [\partial_{t},\partial_s]=0$ for $X$. Then the isometry 
condition  exactly preserves, $\dd s\equiv \dd s_\qcl$ by
the definition (3-3). This isometry relation  should be compared with
 (2-6) which is isometry modulo $\epsilon^2$. It also should be noted 
 that although $\epsilon$ is constant, dependence of the variation 
 upon the position $s$ is given though (3-4) and $u_a(s)$, $a=1,2$. 
 Thus (3-3) is not trivial deformation in general.

I have  the relation,
$$
   - \partial_{t} \exp(\ii \phi_{\qcl}) = \left((u_{1s}-u_2 k_{\qcl})
                    +\ii(u_{2s}+u_1 k) \right)  \exp(\ii \phi_{\qcl}) 
                        . \tag 3-5
$$
Noting that $\phi$ and $u$'s are real valued,
 I obtain (2-7) again from the first term and  solve the differential 
equation between $u_1$ and $u_2$ [8,10], 
$$
	\partial_s u_1 = k_{\qcl} u_2, 
	\quad
       u_1 =\int^s \dd s \ u_2 k_{\qcl} =: \partial_s^{-1} u_2 k_{\qcl} . 
        \tag 3-6
$$
Here $\partial^{-1}_s$ has a parameter $c\in\Bbb R$ as an integral 
constant  and coincides with  the pseudo-differential operator.

Then (3-5) is reduced to the equations,
$$
         \partial_{t} k_{\qcl}=-\Omega u_2   ,   \quad
%
 \Omega:=  \partial_s^2 +\partial_s k_{\qcl}\partial_s^{-1} k_{\qcl} . 
 \tag 3-7
$$
From (3-3) and $ [\partial_{t},\partial_s]=0$ for $X$, $\phi$ is 
calculated as,
$$
	\phi(s,t)=\phi_\qcl(s,t+\epsilon )
	=\ee^{\epsilon \partial_t}\phi_\qcl(s,t)
	 = \phi_\qcl+\epsilon \partial_t \phi_\qcl
	 +\frac{1}{2!}\epsilon^2 \partial_t^2 \phi_\qcl+\cdots .
	 \tag 3-8
$$
Then noting $k^2(s,t)=k_\qcl^2(s,t+\epsilon)$,  the energy functional
 is expressed as
$$
    \split 
  \int k^2 \dd s &= \int\left(k^2_{\qcl} 
  +2 \epsilon  k_{\qcl} \partial_t k_\qcl
  +\epsilon^2((\partial_t k_\qcl)^2+k_\qcl\partial_t^2 k_\qcl)+
       \cdots \right) \dd s
                       \\
   &=\int(k_{\qcl}^2 +2\epsilon k_{\qcl} \Omega  u_2
        +\epsilon^2((\partial_t k_\qcl)^2+k_\qcl\partial_t^2 k_\qcl)+
       \cdots  )\dd s\\ 
       &=:E_{\qcl}+\delta^{(1)} E+\delta^{(2)} E+\cdots
        .\endsplit \tag 3-9
$$
Using (3-6), if I will perform the functional derivative of $E$ in $u_1$, 
I obtain the classical equations (2-12)'s again.
Since the quasi-classical configuration is realized as the extremum point 
of the functional space, I must impose the relation,
$$
	\delta^{(1)} E=0 . \tag 3-10
$$

Noting the relation (3-6), 
if $\Omega u_2$ could be regarded as another function $u_2'$ of the
variation of the normal direction in (3-2), I might find the relation
$$
	\int \dd s k_{\qcl} \Omega  u_2 \sim \int \dd s k_{\qcl}   
	u_2'= \int \dd s \partial_s   u_1' =0
		.\tag 3-11
$$

Accordingly I will introduce the sequence for mathematical times 
$\bold t$ $:=(t_1,t_3,t_5,$$\cdots,t_{2n+1},\cdots )$ so that 
(3-11) is satisfied.  I will redefine the fluctuation (3-2) 
and introduce infinite parameters family,
which can sometimes become finite set as I will show later,
$$
	 X = \ee^{(1/\sqrt{\beta} )\sum_{n=0}\delta t_{2n+1}
	  \partial_{t_{2n+1}}} X_{\qcl}
	 =X_{\qcl} + (1/\sqrt{\beta} )
	 \sum_{n=0} \delta t_{2n+1} \partial_{t_{2n+1}} X_{\qcl}
	  +\Cal O(1/\beta). \tag 3-12
$$
Here  $\epsilon$  was replaced with $(1/\sqrt{\beta} )\delta t_{2n+1}$
and $\partial_{t_{2n+1}} X_{\qcl}$ is expressed as
$$
	\partial_{t_{2n+1}} X_{\qcl}
	     = (u_1^{(n)} + \ii u_2^{(n)} )\exp(\phi_{\qcl}),
	\qquad
        u_1^{(n)} = \partial_s^{-1} k_{\qcl} u_2^{(n)}, \quad
        u_2^{(n)} =  \Omega^n  u_2^{(0)} ,
        \tag 3-13
$$
with integral constants $c$'s vanishing for $n>1$.
 $u_2^{(0)}$ is an appropriate function of  $s$.

Then the variation of the energy functional is calculated as,
$$
    \split 
    \int k^2 \dd s &
         = \int (k_{\qcl}^2 + (2/\sqrt{\beta})\sum_n\delta t_{2n+1}  
             k_{\qcl}\partial{t_{2n+1}} k_{\qcl}) 
    	\dd s
                   +\Cal O (1/\beta) \\
         &=\int(k_{\qcl}^2 +(2/\sqrt{\beta})\sum_n\delta t_{2n+1} k_{\qcl} 
                        \Omega \ u_2^{(n)}) \dd s +\Cal O(1/\beta)\\ 
       &=\int \dd s\ k_{\qcl}^2 +(2/\sqrt{\beta})\sum_n \delta t_{2n+1}
       \int \dd s \partial_s u_1^{(n+1)} +\Cal O(1/\beta) \\
       &=\int \dd s\ k_{\qcl}^2 +\Cal O(1/\beta) 
        .\endsplit \tag 3-14
$$
Thus for the variations along the directions, the energy of the system is 
invariant modulo $1/\beta$.
 Without work, we can move it for these directions $\delta t_{2n+1} 
 \partial_{2n+1} X$ [10,37,38].

However for this sequence, the infinite differential equations appear [8,9],
$$
\partial_{t_{2n+1}} k_{\qcl} =- \Omega^n u_2^{(0)}   .  \tag 3-15
$$
The recursion equations (3-13) are determined by the initial condition 
$u_2^{(0)}$. By following the thought of the quasi-classical method of 
the path integral, this sequence must contain the classical equations 
(2-12)'s. On the other hand, for a close elastica, there is a trivial 
continuous symmetry which is the translation of the 
coordinate system $s$ along the curve $C$. Hence (3-15) should 
also include such translation symmetry.
Of course, they contain other equations as quasi-stable shapes as I 
show following.

Though there might be other choices, I will select minimal subspace of 
the functional space in order to satisfy above requirement. As I 
performed the variational computation in \S 1, I choose initial state as
$$
	u_2^{(0)} =0 , \quad \text{and}
	\quad u_2^{(n)}=\Omega^n \partial_s 
	k_{\qcl}\quad \text{for}\ n\ge 1.   \tag 3-16
$$ 
Then the minimal set of the virtual equations of motion, which satisfies 
the physical requirements, is 
$$
	\partial_{t_{2n+1}} k_{\qcl} =- \Omega^n \partial_s k_{\qcl} ,
	 \quad 
	\partial_{t_{2n+3}} k_{\qcl}
	=\Omega \partial_{t_{2n+1}} k_{\qcl} .\tag 3-17
$$
First several equations and $u$'s are given as follows,
$$
        u_1^{(0)} =1 , \quad u_2^{(0)}=0 ,  \tag 3-18-a
$$ 
$$
      u_1^{(1)} =\frac{1}{2} k_{\qcl}^2 ,
       \quad u_2^{(1)}= \partial_s k_{\qcl} ,  
                      \tag  3-18-b
$$ 
$$ 
    u_1^{(2)} =  \frac{3}{8} k^4_{\qcl} - \frac{1}{2} 
    (\partial_s k_{\qcl})^2 + k \partial_s^2 k_{\qcl} ,
    \quad
   u_2^{(2)} = \frac{3}{2} k^2_{\qcl} \partial_s k_{\qcl} 
             + \partial_s^3 k_{\qcl} ,
         \tag 3-18-c   
$$
$$
     \align
   n= 0:& \quad \partial_{t_1} k_{\qcl} + \partial_s k_{\qcl } =0 , 
                  \tag 3-19-a \\
   n= 1:& \quad \partial_{t_3}k_{\qcl} + \partial_s^3 k_{\qcl} 
               +\frac{3}{2} k_{\qcl}^2
                 \partial_s k_{\qcl} =0 , \tag 3-19-b \\
   n= 2:& \quad \partial_{t_5}k_{\qcl} +
         \partial_s^5 k_{\qcl}+\frac{15}{8} k_{\qcl}^4 
         \partial_s k_{\qcl}
         +\frac{5}{2} (\partial_s k_{\qcl})^3 \\
        & \qquad \qquad
         +\frac{5}{2}k_{\qcl}^2 \partial_s^3 k_{\qcl}
            +10 k_{\qcl} \partial_s k_{\qcl } \partial_s^2 k_{\qcl}=0 .
           \tag 3-19-c   \endalign
$$
Since $\Omega$ is identified with the Gel'fand-Dikii operator for the
MKdV equation, (3-17) is regarded as the MKdV hierarchy 
and $u_1^{(n)}$'s are hamiltonian of the MKdV hierarchy [8-10,39].

Next I will consider the solutions of these equations.
(3-19-a) means the freedom of choice of the origin of $s$ and has only
trivial solution $k(s-t_1)$; $t_1$ is the origin of $s$.
Hence I must pay my attention only upon properties of (3-17) for $n\ge 1$.
Derivative of (2-12-b) can be described as
$$
              c\partial_s k= \Omega \partial_s k. \tag 3-20
$$
By $t_3 = s/c$, this is identified with $n=1$ equation (3-19-b).
Here $\partial_s k$ of the solutions of (2-12-b) is interpreted as the 
eigenvectors of $\Omega$ and $c$ is an eigenvalue of the operator $\Omega$. 
 Thus  (3-17) becomes
$$
	\partial_{t_{2n+1}}  k =- \Omega^n\partial_s k = -c^n \partial_s k .
	 \tag 3-21
$$
Hence the solutions of (2-12) can be solutions of all equations in (3-17).
In fact for a stable solutions $k=$constant, all $u_2^{(n)}\equiv 0$ and 
thus it satisfies all equations in (3-17). 

It should be noted that 
(3-21) comes from the fact that  $u_1^{(m)}$'s agree with the hamiltonians
of the MKdV hierarchy and are regarded as conservation quantities for 
$n$-th equations of $n<m$ [39]. Hence using soliton theory, any 
solutions of (3-19-b) are  solutions of  higher equations in (3-17) 
$n\ge 2$. Due to the requirement of the quasi-classical solutions, 
any solutions must  satisfy all of $n$-th equations ($n\ge1$) in (3-17).
Hence I should deal with only the solutions of the MKdV equation (3-19-b)
as the quasi-classical solutions of this system.

Then the sequence (3-17) fulfills the physical requirements.
In other words,  the solution space of the MKdV equation (3-19-b) is 
the minimal space containing the classical solutions and translation 
symmetry and filling the quasi-stable condition (3-10). 

Since for the variations along the directions $\bold t$ of (3-17), 
the energy of the system (3-14) is invariant,
 the fluctuation of the quasi-classical shape $k_{\qcl}(s,t_{2n+1})$ 
 should be regarded as (generalized)  jacobi-fields or the 
 Goldstone mode [37,38] 
even though they do not obey a linear differential equation in general. 

Here I will remark that the MKdV hierarchy naturally appears by 
physical requirement. It is very surprising because the MKdV 
hierarchy has infinite hamiltonians and
time axes; in classical theory, these quantities cannot be physically 
interpreted. Furthermore it should be contrasted with  the works 
related to space curves in [8,9,33], 
in which the authors chose (3-16) by hand without any physical requirement.

Due to the fluctuation of the heat noise, the equation of 
the elastica in heat bath obeys the MKdV equation (3-19-b) rather 
than the static MKdV equation (2-12).
Let the set of solutions referred as
$$
	\frak S_\qcl=\{C | C\ \text{is a solution of (13-19-b)}\} . 
	\tag 3-22
$$
Of course $\frak S_\qcl$ contains $\frak S_\cls$.
In $\frak S_\qcl$,  various shapes appear as the quasi-classical 
solutions in heat bath. For example, there should exist a deformed 
circle as a solutions of (3-19-b). As another example, there are 
other topological solutions of the different sector of the fundamental 
group,
$$
	 \frac{1}{2 \pi} \int \dd \phi_{\qcl} \in \Bbb Z. \tag 3-23
$$

\subheading{Moduli of Closed Quasi-Classical Elastica}

Here I will go back to compute the partition function (3-1), 
which was formally defined.
First the problem aries how (3-1) should be interpreted.
According to the philosophy of the canonical ensemble, 
the partition function should be calculated by the sum allover 
distinguishable and possible curves satisfying the isometry condition 
with Boltzmann weight.
In the calculus, different topological class of (3-23) will be summed over.

However the partition function (3-1) naturally diverges because the energy
function is  invariant for the affine transformation {\it i.e.} for the
translation and the global rotation.
Fixing $C$, if I denote $X$ as
$$
	X(s) = X_g + X_r(s), \quad \int \dd s X_r(s) =0, \tag 3-24
$$
where $X_g$ is the centroid of the curve $C$, the measure of the
partition function can be rewritten as
$$
	\int D X \ee^{-\beta E} = \sum_{C }
	 \int D X_g \int D X_r \ee^{-\beta E(C)}. \tag 3-25
$$
The $\int D X_g$ is  volume of the base space $\Bbb CP^1$ and diverges.
Including the rotational symmetry and the inner translation symmetry 
$s \to s-t_1$, I will redefine the partition function which
is divided by the volume of the affine transformation of the base space 
and length of the elastica,
$$
 Z_{\reg}:= \frac{\int D X \ee^{-\beta E}}{\text{vol(Aff}(\Bbb CP^1))L}. 
 \tag 3-26
$$

Then I concentrate the shape of the elastica. I must classify the
shape of the elastica and sum over the possible shapes.
In other words, I must investigate the moduli space of 
the quasi-classical elastica, 
$$
	\frak M_\qcl := \frac{\frak S_{\qcl}}{\text{Aff}(\Bbb C P^1) 
	\times S^1} .\tag 3-27
$$

First I will consider the moduli space of the MKdV equation.
The moduli of the MKdV equation was investigated as the KP-hierarchy
using Sato-theory [40-42]. (By the Miura map, the MKdV hierarchy
is transformed to the KdV hierarchy and 
the KdV hierarchy is a subset of the KP hierarchy [41].)
The moduli of the MKdV equation is classified with the genus 
$g \in \Bbb N$ of the hyperelliptic Riemannian surface 
(hyperelliptic curve) $R_g$, which is the finite gap energy 
manifold (Bloch band spectrum) of the wave functions in the inverse 
scattering method of the MKdV equation [40-46].
Hence the moduli of the closed elastica is also classified by the genus,
$$
\frak M_\qcl=\coprod_{g} \frak M_\qcl^{(g)}. \tag 3-28
$$
I will call the genus of the MKdV equation with the boundary condition
genus of the elastica $g$.
In fact, the classical solutions of  (2-12)'s and (3-20) correspond to 
the elasticas of genus zero and one because these energy manifolds 
appearing in its inverse scattering method exhibit a Riemannian sphere 
and an elliptic curve respectively [43].
(It should be noted that even in the quasi-classical equation, 
the solutions of the circle and the eight-figure are
 unique up to homothety for $g=0$ and $g=1$ respectively.)
Using the knowledge of the properties of the universal Grassmannian
manifold (UGM), I will consider the moduli of the closed elastica.

First I will consider the simplest case ($g=0$), a circle $k=2 \pi n/L$,
$n \ge 1$. In (3-9), the quasi-classical action of these circles,
$$
E_\qcl[C_n] = \frac{2 \pi n^2}{L}, \quad k(C_n)=\frac{2 \pi n}{L}
	\tag 3-29
$$
Hence for large $n$, the Boltzmann weight $\exp(-\beta E_\qcl[C_n])$
rapidly decreases. This situation preserves for the elasticas
of higher genus.

Next I will consider elastica of genus one and why the number of the 
genus one solutions of closed elastica is only one, {\it i.e.}, 
eight-figure shape, up to scaling. The moduli of compact Riemannian 
surface of genus one (or elliptic curve) is conventionally expressed 
as $(1,\tau)$; $\tau \in \tilde{\frak M}_{R_1}$;
$$
	\tilde{\frak M}_{R_1}=H_+/\PSL(2,\Bbb Z),
$$ $$
	H_+:=\{ m \in \Bbb C| Im(m)\ge 0\},
$$ $$
 \PSL(2,\Bbb Z) :=\left\{\pmatrix a & b \\ c & d\endpmatrix
	       \Bigm| ad-bc =1, a,b,c,d \in \Bbb Z\right\}/(\pm1). 
	          \tag 3-30
$$
However there is a dilatation freedom 
$(\tilde K,\tilde K':=\tilde K\tau)$ and 
thus I will denote ${\frak M_{R_1}}:=$
$\Bbb R_{>0} \times \tilde{\frak M}_{R_1}$ 
to include the freedom: $\tilde K \in \Bbb R_{>0} :=\{x \in \Bbb R|x>0\}$.
The Jacobi variety of an elliptic curve is given as 
$J_{1,\bold m}:=\Bbb C/(\tilde K\Bbb Z\oplus\tilde K'\Bbb Z)$
for $\bold m:=(\tilde K,\tilde K')$. Since  $\phi_{\qcl}$ is a real analytic 
function of $s \in S^1 = \Bbb R/ L \Bbb Z$, its domain embedded in 
$J_{1,\bold m}$ must be real. Thus  only one-dimensional parameterization of 
$\phi_{\qcl}$, $S^1 \subset J_{1,\bold m}$, is allowed, which is direct line 
in $J_{1,\bold m}$ and passes its origin, because $J_{1,\bold m}$
is complex one-dimensional manifold. 
Since the moduli was divided  by $\text{PSL}(2,\Bbb Z)$,
there are $\PSL(2,\Bbb Z)$ choices how such $S^1$ is embedded 
$J_{1,\bold m}$.  I choose a function with period $L/n$, 
$n \in \Bbb N:=\{n \in \Bbb Z| n\ge 1\}$ 
 $\phi_\qcl(s+L/n)=\phi_\qcl(s)$.
By the periodicity of $\phi_\qcl(s)$, 
I will fix the $\tilde K$ for each embedding of 
$S^1$ into $J_{1,\bold m}$, then the moduli of the MKdV equation with 
period $L$ 
is ${\frak M_{R_1}}\times\Bbb N$
$ \times \PSL(2,\Bbb Z)/\Bbb R_{>0}$, which is equivalent with 
$\Bbb N \times H_+$.

On the other hand, the closed condition (2-19) in $\Bbb CP^1$
restricts the moduli of the elastica. I will introduce a real analytic map,
$$
	f_1 :\dfrac{\frak M_{R_1}\times\Bbb N \times \PSL(2,\Bbb Z) }
	{\Bbb R_{>0}}
	\approx \Bbb N \times H_+\to \Bbb CP^1 ,
$$ 
$$
	    f_1(  m) = X(L)-X(0) . \tag 3-31
$$
Both $H_+$ and $\Bbb CP^1 $ are complex one-dimensional spaces.
The moduli of closed elastica with genus one is given as the inverse image  
of zero point of $f_1$,
$$
	\frak M_\qcl^{(1)} = f_1^{-1}(0). \tag 3-32
$$
Due to the analyticity of the map $f_1$,  $\frak M_\qcl^{(1)}$
is zero-dimensional  manifold. Thus the kind of the shapes are countable
and due to the uniformity, there is only one-solution for each $n \in \Bbb N$.

In ordinary computations [4,10-14,16,32,33], 
by reparameterizing above $S^1$ as $\Bbb R/(L/n) \Bbb Z$,
one starts with $\Bbb C/((L/n)\Bbb Z\oplus K'\Bbb Z)$, $n \in \Bbb N$ and 
$K'\in H_+$ without dividing $\PSL(2,\Bbb Z)$
and searches for the point satisfying the closed condition (2-19).

Similarly properties of the moduli of the closed elastica with genus 
$g\ (>1)$ will be investigated. It is well-known that by the Sato theory,
the characteristic of the KdV hierarchy in the KP hierarchy is to 
characterize its energy manifold in the inverse scattering method 
as  hyperelliptic curve in general (compact) Riemannian surfaces [40-42]. 
The Miura map from the MKdV hierarchy to the KdV hierarchy are bijective. 
Thus I will deal only with the hyperelliptic curves in this article.
First I will denote the moduli of the hyperelliptic curve of $g(>1)$ as 
$\tilde {\frak M}_{R_g}$. Its element is conventionally expressed as 
$(I_g,T_g)$, where $I_g$ and $T_g$ are $g \times g$ matrices;
$T_g=(\pmb{\tau}_1,\cdots,\pmb{\tau}_g)=(\tau_{ij})$ and  
$I_g=(\bold e_1,\cdots,\bold e_g)$ is the unit matrix. 
As I did in $g=1$ case, I will deal with $\tilde K (I_g,T_g)$, 
$\tilde K \in \Bbb R_{>0}$ rather than $(I_g,T_g)$ itself.
It is known that the dimension of the moduli of the hyperelliptic curves,
$\tilde {\frak M}_{R_g}$, is $2g-1$. Then I will also introduce a real 
$2g$-dimensional lattice for a point of the moduli 
$ \bold m \in \frak M_{R_g}:=\Bbb R_{>0}\times \tilde {\frak M}_{R_g}$ 
[44-46],
$$
	\Gamma_{\bold m} = \left
	\{\sum_{j=1}^g m_j \tilde K \bold e_j 
	+\sum_{j=1}^g n_j \tilde K \pmb{\tau}_j
	\Bigm|(m_i,m_j \in \Bbb Z)\right\} ,
	\tag 3-33
$$
and the Jacobi variety $ J_{g,\bold m}:=\Bbb C^g/\Gamma_{\bold m}$.
If I determine a point $\bold m$ of $\frak M_{R_g}$, I can uniquely 
construct the Jacobi variety, $ J_{g,\bold m}:=\Bbb C^g/\Gamma_{\bold m}$.
From the soliton theory, if the coordinates of $J_{g,\bold m}$ as 
real manifold are expressed by $\bold t_{KP}=(t_1,t_2,t_3,t_4,\cdots,
t_{2g})$ [40-42], its subset with odd indices 
$\bold t_g:=(t_1,t_3,\cdots,t_{2g-1})$ can be identified with the 
part of $\bold t$ in (3-12). This identification can be 
guaranteed by the Krichever construction of the solution of the 
KP hierarchy [44]  and Sato theory [40-42]. By the  Krichever 
construction, it is known that  each 
parameterization $t_n \in \bold t_{KP}$ is direct line passing 
the origin in the Jacobi variety. 
(3-17) and (3-19) are reduced to the linear differential equations in the 
Jacobi variety [44-46]. (Thus (3-17) can be recognized as 
the Jacobi equation of the Jacobi-field of the system (3-14).)

Since the moduli of the hyperelliptic curves has been also divided 
by a discrete group $\SSp(g,\Bbb Z)$ [44] like $\PSL(2,\Bbb Z)$ of 
$g=1$ case, for fixing $\bold m \in  \frak M_{R_g}$, there are 
$\Bbb N \times\SSp(g,\Bbb Z)$ 
ways to embed $S^1$, as a period of $\phi$, into $J_{g,\bold m}$. 
The number of the ways  are equivalent to cardinal of 
$\Bbb N \times\SSp(g,\Bbb Z)$ as a set.
As I impose its periodicity of $L/n$ ($n \in \Bbb N$) on $S^1$, the dilation 
parameter $\tilde K$ is determined. 
Let such moduli space denoted as 
$$
      \frak M_{R_g,S^1}:=\frak M_{R_g} \times 
      \Bbb N\times \SSp(g,\Bbb Z)/\Bbb R_{>0}. 
      \tag 3-34 
$$
By choosing a point of the moduli space $\frak M_{R_g,S^1}$,
a  $g \times g$ lattice $\Gamma_{\bold m,S^1}$ 
is uniquely determined and Jacobi variety  is given as,
$$
       	\frak M_{R_g,S^1} \to \{\Gamma_{\bold m,S^1}\}, 
       	\quad J_{g,\bold m,S^1}:=\Bbb C^g/\Gamma_{\bold m,S^1}
       	. \tag 3-35
$$

I will fix a point of the modulus $\bold m \in \frak M_{R_g,S^1}$ for a while.
From the properties of the MKdV hierarchy, $\phi_{\qcl}$ is a real 
analytic function of  $\bold t_g$. As I can expand it around a point 
$\bold t_g$ using the properties of its real analyticity,
$$
	\phi_{\qcl}(\bold t_g')
	= \sum_{n_0,\cdots,n_g} a_{n_0,\cdots,n_1}(t'_1-t_1)^{n_0}
	(t_3'-t_3)^{n_1}\cdots(t_{2g-1}'-t_{2g-1})^{n_g} ,
	\tag 3-36
$$
$\bold t_g$ must be a system of real parameters in $ J_{g,\bold m,S^1}$. 
By analytic continuation, $\bold t_g$ can be locally complexfied.
On the other hand, the Jacobi variety has a canonical complex structure,
$$
	 \Cal J: J_{g,\bold m,S^1} \to  J_{g,\bold m,S^1} , \quad \Cal J^2=-1 
,
			\tag 3-37
$$
which consists with its affine (vector) structure.
By the structure $\Cal J$, there are set of the real 
$g$-dimensional submanifolds 
$\{\Sigma_{g,\bold m,S^1}\}$ which includes the orbit of $s(\in S^1)$ as 
its one-dimensional submanifold. Due to the analyticity of $\phi_{\qcl}$ over 
$J_{g,\bold m,S^1}$, these complexfication of (3-36) cannot  contradict 
with this complex structure $ \Cal J$. Then $\bold t_g$ can be regarded as an
element of $\Sigma_{g,\bold m,S^1} $. I will refer such embedding,
$$
	\sigma_0 : \Sigma_{g,\bold m,S^1}  
	\hookrightarrow  J_{g,\bold m,S^1} ,  \tag 3-38
$$
and the set of $\sigma_0$ is expressed as $\Em_0(\Sigma_{g,\bold m,S^1},  
J_{g,\bold m,S^1})$. Since $(t_3,\cdots,t_{2g-1})$ need not be periodic, 
this embedding is measurable and $\Em_0(\Sigma_{g,\bold m,S^1}$,  
$J_{g,\bold m,S^1})$ can be regarded as a subset of 
the Grassmannian manifold, 
$\Em_0(\Sigma_{g,\bold m,S^1},J_{g,\bold m,S^1}) \subset $
$\GL(\Bbb R,2g-1)$$/\GL(\Bbb R,g)\GL(\Bbb R,g-1)$.

Then I can construct the fiber structure for 
$\Em_0(\Sigma_{g,\bold m,S^1},  J_{g,\bold m})$,
 because for a way to such embedding, there is the trajectory space
$ \Sigma_{g,\bold m,S^1}/S^1$ as fiber space, where
$(t_3,\cdots,t_{2g-1})$ $\in$ $ \Sigma_{g,\bold m,S^1}/S^1$.
I refer the fiber bundle as $\frak F_{\bold m}$,
$$
      \pi_{\traj}:\frak F_{\bold m} \to \Em_0(\Sigma_{g,\bold m,S^1}, 
       J_{g,\bold m,S^1}) .  \tag 3-39
$$
This fiber space also depends upon the point of the moduli space of the 
Riemannian surfaces $\frak M_{R_g,S^1}$.
Hence the moduli of the periodic solutions of $\phi_{\qcl}(s,\bold t_g)$, 
which is written as ${\frak M}_{\period}^{(g)}$, has also fiber structure,
$$
	\pi_{\period} :
	{\frak M}_{\period}^{(g)}\to \frak M_{R_g,S^1}. \tag 3-40
$$
For each point $ \bold m \in \frak M_{R_g,S^1}$, 
the fiber bundle $\frak F_{\bold m}$ (3-39) stands up as a fiber of 
${\frak M}_{\period}^{(g)}$.

By the closed condition, I must restrict the moduli space.
Here I will consider a real analytic map like (3-31),
$$
	 f_g :{\frak M}_{\period}^{(g)} \to \Bbb CP^1,
$$
$$
	    f_g( \mu  ) = X(L)-X(0) .
                 \tag 3-41
$$

Consequently I obtain the moduli of the closed elastica, 
which is expressed as 
$$
     {\frak M}_\qcl^{(g)}= f_g^{-1}(0) \subset {\frak M}_{\period}^{(g)} .
	\tag 3-42
$$
Since  image of $f_g$ is real two-dimensional manifold,
for $g>2$, dim$({\frak M}_\qcl^{(g)})\ge 1$ and ${\frak M}_\qcl^{(g)}$
is measurable.

For simplicity, I will introduce the notations:
$$
	{\frak M}^{(g)}_{\bold t_m=0}:=\pi_{\traj}( {\frak M}_\qcl^{(g)}), 
$$
$$
      {\frak X}^{(g)}\ni n:{\frak M}^{(g)}_{\bold t_m=0} 
      \to {\frak M}^{(g)}_{\bold t_m=0,n},
      \quad \text{or}\quad {\frak M}^{(g)}_{\bold t_m=0}
      =\oplus_{n \in {\frak X}^{(g)}}{\frak M}^{(g)}_{\bold t_m=0,n}
$$ $$
       \text{for}\  (n,\frak m) \in {\frak M}^{(g)}_{\bold t_m=0},\quad
	\tilde \Sigma_{\frak m,n}:= \pi_{\traj}^{-1}(n,\frak m)
	 \tag 3-43
$$
where ${\frak X}^{(g)}$ is countable part of ${\frak M}^{(g)}_{\bold t_m=0}$ 
and  ${\frak M}^{(g)}_{\bold t_m=0,n}$, the restriction of 
${\frak M}^{(g)}_{\bold t_m=0}$ for a point $n \in {\frak X}^{(g)}$, 
is measurable part. 
Here $\tilde \Sigma_{\frak m,n}$ has coordinate $(t_3,\cdots,t_{2g-1})$ $
=:\bold t_{\frak m}$.
Hence there is a map from the moduli to the shape of the elastica.
$$
       h: {\frak M}_\qcl^{(g)} \ni (n,\frak m,\bold t_{\frak m})
      \mapsto C_{\frak m}^{(n)} (\bold t_{\frak m}^{(n)}) \subset \Bbb C P^1.
       \tag 3-44
$$

For a perturbative deformation like distortion to an ellipse from a circle, 
the energy manifold in the inverse scattering method has infinite gaps 
(or genera) in general. Thus such deformation is expressed in the moduli 
of $g \to \infty$ and due to the integrability of the MKdV equation, the 
deformation can be predicted like harmonic oscillator around a stable point.
This picture is supported by the linearized method of the nonlinear equation
and is also built in the above argument of the limit $g \to \infty$.

\subheading{Partition Function}

As I finish to classify the solution space formally, I will consider
the fluctuation of the elastica again.
It should be noted that there is above limit of  the sequence $u_2^{(n)}$
corresponding to the genus  of the elastica. 
If I encounter for $n=N$
$$
	\Omega u_2^{(N)}\equiv \lambda u_2^{(N)},\quad  \lambda\in \Bbb R, 
	\tag 4-45
$$
like (3-20), then $\delta t_{2(N+m)+1}\propto \delta t_{2N+ 1}$ for $m>0$
  because of (3-12) and
$$
	\partial_{t_{2(N+m)+1}} u_1^{(N+m)}=k_{\qcl}  u_2^{(N+m)}=k_{\qcl}
	 \Omega^m u_2^{(N)}
	=\lambda^m \partial_{t_{2N+ 1}} u_1^{(N)} . \tag 3-47
$$
Accordingly there needs no other fluctuation parameter $n>N$ because these
fluctuation vectors are linearly dependent.  The sequence 
of (3-9) should be truncated according to the  philosophy of the canonical
 ensemble.
 Thus I will denote such minimal integer, 
which is a function of the solution, as
$$
	\text{ind}_0 : C \to N(C) \in \Bbb Z. \tag 3-48
$$
However from the soliton theory [40-42] and above argument, 
for $C \in \frak M_\qcl^{(g)}$, I conclude that $\text{ind}_0(C)=g$.
Avoiding meaningless divergence, I will replace the infinite series in 
(3-12)
with finite sum from $1$ to $g$ depending upon the shape of elastica.

Since the direction of $\delta t_1$ is along the tangential 
direction of the elastica $C_\qcl$, its effect has been treated as 
the integral of $\delta t_1\propto s$ in (3-26). 
On the other hand, $\delta t_{2n+1}$ ($n>1$)
includes the normal direction fluctuation and I must integrate
the Boltzmann weight over $\delta t_{2n+1}$ space depending upon the 
genus of the elastica.
Linear independence of these bases are guaranteed by above truncation.

Then for a curve $C \in \frak M_\qcl^{(g)}$, the heat fluctuation of 
higher order is  expressed as
$$
	\delta^{(n)} E[C,\delta  t_{2m+1}]=\sum_{0<m_i\le g}\frac{1}
	{\sqrt{\beta}^n} \prod_i^{n}
	 \left(\delta t_{2m_i+1}\right) \int \dd s \prod_i^{n}
	 \left( \partial_{t_{2m_i+1}}\right) k^2
	 .\tag 3-49
$$
Here $m_i=0$ part vanishes due to the periodicity 
$$
	\int \dd s \partial_s\left( \prod_i^{n-1} 
	\left(\partial_{t_{2m_i+1}} \right)k^2\right)=0. \tag 3-50
$$
On the other hand, if the set \{$m_i$\} does not contain $m_i=0$ component,
the integral commutes with these derivatives,
$$
     \int \dd s\ \prod_i^{n} \left(\partial_{t_{2m_i+1}} \right)  k^2
     = \prod_i^{n} \left(\partial_{t_{2m_i+1}} \right)\int \dd s  \ k^2. 
     \tag 3-51
$$
Since $\int k^2 \dd s$ is invariant for the time $t_{2n+1}$ ($n>0$) 
development from the soliton theory, (3-51) vanishes. Hence 
I obtain that all higher order fluctuations   vanish,
$$
	\delta^{(n)} E[C,\delta  t_{2m+1}]\equiv 0, 
	\quad \text{for} \ n>0 . \tag 3-52
$$

In other words, the effect of heat fluctuation is given only through 
the energy functional of the quasi-classical motions.
Since for a quasi-motion of genus $g$, the curvature is precisely given as
$$
	k(s, t_1,t_2,\cdots,t_{2g-1})=
	k_\qcl(s, t_1+\frac{1}{\sqrt{\beta}} \delta t_1,
	t_2+\frac{1}{\sqrt{\beta}} \delta t_2,\cdots,t_{2g-1}
	\frac{1}{\sqrt{\beta}} \delta t_{2g-1})
		, \tag 3-53
$$
I must integrate the Boltzmann weight $\exp(-\beta \int \dd s k^2)$ 
over all $\delta t$'s except $\delta t_1$.
Using the translation symmetry and freedom of the integration variable, 
I regard that $t_{2n+1}=\delta t_{2n+1}/\sqrt{\beta}$.

Consequently I obtain an explicit form of 
the regularized partition function (3-26), which is expressed by
$$
	\split 
	Z_{\reg}[\beta] & = \sum_g\sum_{C \in \frak M_\qcl^{(g)} }
	\left(\exp(-\beta E_{\qcl}[C]) \right)\\
	   &= \sum_{g=0}^1 \sum_{n \in \frak X_g^{0}}
	    \left(\exp(-\beta E_{\qcl}[C^{n}]) \right) \\
	   &\qquad +\sum_{g=2}^\infty\sum_{n \in {\frak X}^{(g)}}
	   \int_{{\frak M}^{(g)}_{\bold t_m=0,n}} \dd \frak m 
	   \int_{\tilde \Sigma_{\frak m,n}} \left(\prod_{n=2}^g
	   \dd t_{2 n-1} \right)
	    \exp\left(-\beta E_{\qcl}[C^{(n)}_{\frak m}
	    (\bold t_{\frak m}^{(n)})]
	   \right) 
	    .\endsplit \tag 3-54
$$
This is the exact form of the partition function (3-1) of the 
non-stretched elastica without divergence.
In the second term, there appear the integration of the type of 
$\int \dd x \ee^{-\beta f(x)}$.
Thus it is expected that the prefactor of the second term begins 
with the negative power of $\beta$.
For large $\beta$, the second term is less than the first term.
Hence for the zero temperature limit $\beta \to \infty$, 
the second term disappears
and only the contribution of the genus $0$ and $1$ survives. 
Noting that the moduli of the quasi-classical elastica with
$g\le 1$ is equivalent with the
that of the classical, I obtain
$$
	\split
	\lim_{\beta \to \infty}Z_{\reg}[\beta]
	&= \max_{C \in\frak S_{\cls}}
	  \exp(-\beta E_{\cls}[C])\\
	  &=\exp(-\beta \min_{C \in\frak S_{\cls}}E_{\cls}[C]). 
	  \endsplit\tag 3-55
$$
Depending upon the boundary condition, the classical solutions appear
as minimal points of the partition function $Z_{\reg}[\beta]$.
Hence this partition function (3-54) does not contradict with 
the discovery of Daniel Bernoulli [1]. 

\tvskip
\centerline{\twobf \S 4. Discussion}
\tvskip

It is worth while noting that due to the isometric condition, 
I have derived the  MKdV hierarchy. In the elastic body, the lagrangian 
coordinate system should be employed rather than the
Eulerian coordinate system when I will use the terms of the fluid 
mechanics.
In the elastic body theory by marking some points on an elastic 
body and by estimating variation of distance among the marking points
which is measured using the induced metric, the force will be locally 
evaluated 
as linear response for its certain deformation. 
The marking points corresponds 
to the Lagrangian test particle in the language of fluid mechanics. 
On the other hand, as I have used the metric induced from the base space
and calculated the deformation, my calculation corresponds to 
the Eulerian one.
Here it should be noted that  if one uses the induced metric or Eulerian 
picture, any stretching (physical) curve can be regarded as a non-stretching 
(mathematical) curve; it is a trivial trick between the lagrangian picture 
and Eulerian picture and such recognition has few physical meanings.
If stretching has physical meaning like an elastic body, 
Eulerian picture does not exhibit dynamical situation and unless stretch 
plays important role like boundary curve of binary fluid,  
dealing with stretch has less physical meaning.
Accordingly the isometric condition I employed plays the central role in this
scheme. In other words, in above computation, the reason why I could 
physically use the Eulerian picture even in the elastic body problem is 
owing to this isometry condition.

It should be also noted that even though there appears non-linear
differential equation in this scheme, I have used the energy functional
which is locally given in the framework of linear response of the force
for the deformation [4,11]; if one uses the non-linear energy functional,
he must evaluate it from basic elastic body theory because it is beyond
the ordinary elastic body theory. It is remarked that due to the bilinearity 
of the energy functional, which is established in the framework of the 
ordinary elastic body theory, I could find the exact partition function 
(3-54) in this model.

Furthermore the origin of the MKdV hierarchy in ref.[8,9,33]
was artificial and was not physically supported. If one physically sets up
problem of time development of the elastica for real physical time, 
he concludes that its motion is not governed by
the MKdV equation nor the MKdV hierarchy in general [10-14].
However in this article, I obtained the MKdV hierarchy from the physical 
requirement and a (mathematical) parameter time $\delta t_{2n+1}$ appears
 variational direction as I pointed out in ref.[10].
In other words, by virtue of the novel investigation of  the properties 
of isometric curve of Goldstein and Petrich [8,9], I conclude that
the virtual dynamics is realized as thermal fluctuation of an elastica 
in heat bath. Due to the isometry condition, 
these equations become non-linear differential equations.
In the linear differential equation such as the harmonic oscillator,
the mode, which is determined by the global feature of the system, 
is represented by a vector of momentum space. As well as mode analysis 
of the linear system, these parameters $t_{2n+1}$ exhibit the global 
deformation of the elastica due to the thermal fluctuation
and is expressed as a vector in the Jacobi variety.

It is remarked that  the obtained partition function (3-54) differs from 
that in ref.[24], which is obtained by summing the weight function over 
the conformation including non-isometry deformation. 
Due to the isometry condition, non-linear terms appears
in the quasi-classical curve equation while the partition function 
proposed by  Sait\^o {\it et al.} [24] is essentially linear. 
However for perturbative deformation, {\it e.g.} from the circle, 
the non-linear term might be negligible. Thus  as long as the deformation 
is in perturbative, their partition function can be applicable even 
for a polymer which cannot be stretched even in thermal fluctuation.

On the other hand, my partition function is justifiable even for large 
deformation. The partition function is summed over different topology $g$, 
which is related to the writhing number of the conformation. Hence 
there is a possibility of the topology change due to the thermal fluctuation.
It is of interesting to calculate the possibility 
(or kernel function) from $g=1$ conformation to $g=2$ conformation.
Even though the partition function (3-54) has not been concretely calculated,
such computation, in principle, can be performed.

Next I will comment on the physical meaning of $\delta t_{2n+1}$ 
and the relation 
between the BRS transformation [31] and the Sato coordinate [40-42].
Since I have dealt with the SO(2) principal bundle over $S^1$, the 
gauge group is expressed as 
$$
	\frak G\subset \coprod_{s\in S^1} \text{SO}(2), \tag 4-1
$$
where $\coprod$ means the disjoint union.
$\frak G$ is infinite dimensional Lie group. It acts upon the shape of 
the curve, which corresponds to a section of the principal bundle, 
and deforms it. For a given shape of elastica, there is a 
unique group element which acts the elastica to become the shape
 with constant curvature, {\it i.e.},
the simplest classical solution with $g=0$. 
Thus the genus is well-defined, which is induced from the genus of curve 
(the quasi-classical section).  There is a decomposition of $\frak G$ 
as a family of subgroups $\frak G_g$ respect to the genus, whose action on
the elastica preserves its genus.
The representation of each group  $\frak G_g$
 will be realized as $\frak G_g$ module in the set of 
 corresponding Jacobi varieties.
However in the soliton theory, instead of 
dealing with  individual sets of Jacobi varieties of genus $g$,
it is natural to consider the UGM if one wishes to
formally treat a soliton equation. In fact there are singular elements
in $\frak G$, which change the genus of the elastica; the 
transformation are known as the global gauge transformation.
Corresponding to UGM, $\frak G$ should also be regarded as 
inductive limit of the
filteration of  $\frak G_g$ respect to the genus $g$ and then 
$\frak G$ naturally
contains the singular elements due to the natural extension of t
he group action.
Thus $\frak G$ is represented as a subset of  GL$(\infty)$ in the UGM.
The quasi-classical curve (a section of the SO(2) principal bundle)
is embedded in the UGM. The infinitesimal deformation of the curve in 
the UGM can  be expressed by a vector in the UGM.
In other words, such deformation exhibits (mathematical) velocity of a 
trajectory in the  UGM
and can be represented as subset of the infinite dimensional general 
linear Lie algebra gl$(\infty)$, which is known as the affine 
Lie algebra $A^{(1)}_1$ [41];
$A^{(1)}_1$ is the Lie algebra associated with the Lie group $\frak G$.
I will introduce the extrinsic differential operator in the UGM
$$
	\delta := \sum_n \dd t_{2n+1} \partial_{t_{2n+1}}, 
	\quad \delta^2=0. \tag 4-2
$$
Then (3-15) and (3-17) are expressed as for $A:=k_{\qcl}\dd s$,
$$
         \delta A=\tilde \Omega \frak u_1, \tag 4-3
$$
where
$$
	\frak u_1=\sum_n u_1^{(n)} \dd t_{2n+1},
	 \quad \tilde \Omega:=\Omega k_{\qcl}^{-1} \partial_s. \tag 4-4
$$
Noting the fact that $u_1^{(n)}$ is hamiltonian density of the MKdV 
hierarchy,
$$
	\delta \frak u_1 \approx 0 , \tag 4-5
$$
where $\approx$ means equivalence after integration the both sides over 
$s$ like (3-50) and (3-51).
Since $\frak u_1$ obeys the Grasmannian algebra, 
$\frak u_1$ can be regarded as fermionic field over $S^1$. 
Consequently (4-3) and (4-5) can be regarded as the BRS transformation 
of this system. Hence  $\delta t_{2n+1}$ in the path integration may 
be naturally  understood in the framework of the Faddeev-Popov 
integration scheme [31].
In fact the square root of the Frenet-Serret system (2-2) can be regarded 
as the Dirac operator [11,12], which is realized by confining the free 
Dirac field into a thin elastic rod. Using the Dirac field confined 
in the elastica, I  constructed the MKdV hierarchy
and $\tau$-function as the partition function of the Dirac field [47]. 
Thus I expect that the  partition function (3-54)
should also be expressed by the $\tau$-function of the MKdV hierarchy.

Here I will mention a conjecture associated with the critical 
phenomenon of this elastica model.
The critical point must be determined as a topological 
discontinuity of the moduli space 
of the quasi-classical elastica. At the point, physical 
quantity sometimes diverges and becomes meaningless. 
 I expect that the length of the elastica  becomes less important 
 around the critical point, {\it e.g.} the topological change, 
 as a kernel function at an ordinary second 
order critical point becomes scale-invariant [48].
I will consider a dilatation of a quasi-classical elastica for 
the normal direction of elastica,
$$
	X_c=X + \ii t \ee^{\ii \phi_c}, \tag 4-6
$$
where $X_c$ and $\phi_c$ are an affine vector and tangential
 phase of the quasi-classical elastica at the 
critical point and $t$ is a real deformation parameter. 
The infinitesimal  length of the elastica $\dd s_c$ becomes
$$
  \dd s_c = \sqrt{\dd X_c \dd \bar X_c}\approx(1 - k_c t) \dd s, 
  \quad k_c := \partial_s \phi_c,  \tag 4-7
$$
and then the length of elastica is
$$
	\int \dd s_c \approx L - t (\phi_c(L)-\phi_c(0)) . \tag 4-8
$$
Noting that $\phi_c(L)-\phi_c(0)=2\pi n$, $n\in \Bbb Z$, 
this deformation makes the length of the elastica change. 
However this deformation must be compatible with the isometric condition
since I have been dealing with the non-stretching elastica.
Both requires seem to contradict with but the critical point 
is an irregular point at which the contradicted objects coexist.
From (3-2), $\ii t$ in (4-6) must be proportional to 
$u_c :=u_1+\ii u_2$ with (3-18-b), and $u_1$ satisfies  (3-19-b) 
in respect of the deformation parameter $t$.
$$
	 \ii c t = u_c = \frac{1}{4} k_c^2 -
	 \frac{1}{2}\partial_s k_c,  \tag 4-9
$$
where $c$ is a real proprtional constant. Since this relation is 
the Miura map, from (3-19-b), $u_c$ obeys the KdV equation 
$$
            \ii c= \partial_t u_c = 6 u_c \partial_s u_c 
            + \partial_c^3 u_c . \tag 4-10
$$
By integrating it in $s$,  (4-10) becomes
$$
	 \ii c s = 3 u_c^2 + \partial_s^2 u_c . \tag 4-11
$$
For $z:=\ii s$ $w:=u_c/2$ and $c=-2$, (4-11) can be rewritten as
$$
	\partial_z^2 w = 6 w^2 +z.  \tag 4-12
$$
This is  the Painlev\'e equation of the first kind.
Thus I will conjecture that at a critical point of the elastica, 
an expectation value obeys the Painlev\'e equation of the first
 kind [49]. In our scheme, I can formally obtain the series of 
 the ordinary differential 
equations related to the KdV hierarchy (from the Miura map) and 
the Painlev\'e transcendents. Thus the muduli space must be closely 
related to the quantum two-dimesnional 
gravity, in which (4-12) and the KdV hierarchy naturally appear [50-52].
In fact, the loop soliton partially appears in the immersed  
surface in three dimensional space $\Bbb R^3$  [38,53].
By the fermionic study [47,53-57], the immersed surface system
is interpreted as a natural extension of the elastica system and also 
that of the Liouville surface system whose quatum version is known as 
the quantum two-dimesnional gravity [55-56]. Thus I plane to 
investigate the immersed
surface system and reveal the relation between the elastica system 
and the quantum two-dimesnional gravity.
(Here it should be noted that elastica problem is not directly related to 
the string problem in the string theory because the action of the elastica
is biharmonic for $X$ while that of string is harmonic [55].
In non-relativistic space, thickness is more important and biharmonic
equation is very natural.)

As I dealt with the kinetic properties of the large closed polymer and
investigated the moduli of the MKdV equation in $\Bbb C P^1$,
recently another statistical mechanics model of a 
large polymer was reported [58].
Partition function of non-contractible self-avoiding 
two-dimensional polymers in the topological torus,
$$
	T =\Bbb C /(L_1 \Bbb Z + L_2 \Bbb Z) , \tag 4-13
$$
was studied associated with the MKdV equation. 
The partition function of such polymers was also solved by the MKdV equation 
[58] and  is  recognized as its $\tau$-function [59]. 
As I assumed the base space as $\Bbb CP^1$, I can also replace
it with the topological torus. Then the ratio $L:L_1:L_2$ 
becomes important as the boundary conditions but I can formally calculate it. 
Then obtained partition function may be also closely related to the 
$\tau$-function of the MKdV equation. It is a very interesting fact that 
the MKdV equation appears and plays central roles in both theories 
even though they are not directly concerned with as  models.  

Finally I will mention farther possibilities and development of this theory.
I investigated the elastica in two-dimensional space but  
can extend my theory to that in three-dimensional space if I can classify 
solutions of quasi-classical curves of elastica in three-dimensional space. 
Thus I need investigation on the moduli of quasi-classical elastica 
in three-dimensional space including topological properties like a 
knot invariance. Furthermore even though I formally classified 
the moduli of the  quasi-classical elastica
in two-dimensional space, I cannot explicitly draw the shape of 
the closed elastica of $g>1$ now.
Accordingly it is very important for this study to find explicit 
shapes of closed elastica of $g>1$.
If I could explicitly draw any shapes of closed quasi-classical 
elastica in plane of $g=2$ or $3$
using hyperelliptic functions, 
I think, they have enormous effects on this problem.
Since  the summation of the partition function might be converges 
on the genus $g$, to determine the shapes even for small $g$ means 
important steps on this problem. 
Moreover, it is also of interests  to deal with an elastic rod 
which can be stretched as more physical model [14] and higher 
dimensional objects, such as an immersed surface [38,53,56,57].


\tvskip
\centerline{\twobf Acknowledgment}
\tvskip

I would like to thank Prof. S. Saito, Prof. K.~Sogo  and 
Prof.~T.~Tokihiro for helpful discussions and continuous encouragement. 
I am grateful to  Y. \^Onishi for crucial discussions of the
moduli of the elastica and for critial reading of parts of 
this manuscript. I also thank   Dr.~S.~Ishiwata and Dr.~K.~Nishinari
for critical discussions and for interesting experiments of elastica.
I am indebted to Dr. H.~Tsuru for drawing my attention to DNA problem. 
I also acknowledge that the seminars on
differential geometry, topology, knot theory and group theory with
Prof. K.~Tamano and W.~Kawase influenced this work.


\enddocument